\documentclass[prc,twocolumn,showpacs,floatfix,aps,10pt]{revtex4-1}

\usepackage{amsmath}
\usepackage{graphicx}

\begin{document}

\title{
Three-body calculations of triple-alpha reaction
}
\author{S.\ Ishikawa}  \email[E-mail:]{ishikawa@hosei.ac.jp}
\affiliation{
Science Research Center, Hosei University, 2-17-1 Fujimi, Chiyoda, Tokyo 102-8160, Japan
} 

\date{\today}

\begin{abstract}
Recently, the triple-$\alpha$ (3$\alpha$) process, by which three ${}^4$He nuclei are fused into a $^{12}$C nucleus in stars, was studied by different methods in solving the quantum mechanical three-body problem. 
Their results of the thermonuclear reaction rate for the process differ by several orders at low stellar temperatures of $10^7 - 10^8$ K.
In this paper, we will present calculations of  the 3$\alpha$ process by a modified Faddeev three-body formalism in which the long-range effects of Coulomb interactions are accommodated.
The reaction rate of the process is calculated via an inverse process, three-alpha (3-$\alpha$) photodisintegration of a ${}^{12}$C nucleus.    
Calculated  reaction rate is about $10$ times larger than that of the Nuclear Astrophysics Compilation of Reaction Rates (NACRE) at $10^7$ K, and is remarkably smaller than the results of the recent three-body calculations.
We will discuss a possible reason of the difference.
\end{abstract}



\maketitle

\section{Introduction}

The thermonuclear reaction rate of the $3\alpha$ process is known to be the important input to studies of the stellar nucleosynthesis and the stellar evolution (see, e.g., Refs. \cite{Tu09,Fy05}). 
This process at stellar temperature as high as $10^9$ K (resonant region) is dominated by the sequential process in which successive formations of the 2-$\alpha$ resonant state, e.g., ${}^{8}$Be$(0_1^+)$, and then the 3-$\alpha$ resonant state, e.g.,  ${}^{12}$C$(0_2^+)$ (the Hoyle state) play the essential role \cite{Sa52,Ho54}. 
On the other hand, at lower temperatures as $10^7$ K, where kinematical energies of $\alpha$ particles are not high enough to produce the ${}^{8}$Be$(0_1^+)$ resonance as a door way state, the process is non-resonant, and should be considered as a direct three-body reaction. 
The NACRE 3$\alpha$ reaction rate \cite{An99}  is evaluated adapting the sequential process with extensions of  the resonance formula to low energies assuming energy dependent widths \cite{No85,Ga11} as a simulation of the direct reaction.

Because of recent developments in solving Schr\"odinger equations for three-body continuum states
numerically,  there appeared  some three-body calculations of the 3$\alpha$ reaction rate. 
Ogata {\it et al.} \cite{Og09} have first calculated the 3$\alpha$ reaction rate with solving  3-$\alpha$ Schr\"odinger equations by  the method of continuum-discretized coupled-channel (CDCC), in which a three-body wave function is expanded by a set of discretized $\alpha$-$\alpha$ scattering states (Hereafter their rate is referred to as OKK rate). 
Due to huge differences from the NACRE rate at the low temperatures (see Fig. \ref{fig:aaa-ratio} below),  the OKK rate was reported to cause tremendous effects on the stellar evolutionary phenomena \cite{Do09,Pe10,Mo10,Sa10,Ma11,Su11,Ki12}. 
Recently,  calculations in which  3-$\alpha$ continuum states are treated by the hyperspherical harmonics method combined with the R-matrix method, was performed in Refs.\  \cite{Ng11,Ng12} (HHR rate).
In Refs.\ \cite{Is13,Is12}, the present author reported some results of the 3$\alpha$ reaction rate calculated by the Faddeev three-body formalism \cite{Fa61}  modified so that effects of the long-range Coulomb interactions are accommodated, which has been successfully applied to the study three-nucleon scattering systems \cite{Is03,Is09}.
More recently, a method of imaginary-time \cite{Ya12} has been applied to calculate the 3$\alpha$ reaction rate \cite{Ya12b}. 
While these different calculations agree with each other and with the NACRE rate at the resonant region, they differ considerably  at lower temperatures (see Fig. \ref{fig:aaa-ratio} below).

This paper will describe some details of the calculations of the $3\alpha$ reaction rate partially reported in Refs.\ \cite{Is13,Is12}, and will discuss the differences among the calculations.
In the following, after describing a formalism to calculate the reaction rate shortly in Sec. \ref{sec:formalism}, 
results of calculations will be presented in Sec. \ref{sec:calculations}.
In Sec. \ref{sec:discussion}, to understand differences between the present calculations and the others,  CDCC calculations will be performed.
A summary will be  given in Sec. \ref{sec:summary}.

\section{Formalism}
\label{sec:formalism}

\subsection{Basic formalism}

We consider  a system of three $\alpha$ particles 1, 2, and 3, and use Jacobi coordinates $\{\boldsymbol{x}_i, \boldsymbol{y}_i\}$ to describe the three-body system defined as 
\begin{eqnarray}
\boldsymbol{x}_i  &=&  \boldsymbol{r}_j - \boldsymbol{r}_k, 
\cr
\boldsymbol{y}_i  &=&  \boldsymbol{r}_i 
   - \frac12 \left(\boldsymbol{r}_j + \boldsymbol{r}_k \right),
\label{eq:Jacobi}
\end{eqnarray}
where $(i,j,k)$ denotes $(1,2,3)$ or its cyclic permutations, and $\boldsymbol{r}_i$ is the position vector of the particle $i$. 
Momenta conjugate to $\boldsymbol{x}_i$  and $\boldsymbol{y}_i$ are denoted by $\boldsymbol{q}_i$ and $\boldsymbol{p}_i$, respectively.
Subscripts to indicate particles will be omitted when there is no confusion.

Let us consider the electric quadrupole (E2) transition from a 3-$\alpha$ continuum state of the total angular momentum 0  to the  ${}^{12}$C$(2_1^+)$ bound state emitting a photon of the energy
\begin{equation}
E_\gamma = E - E_b = E+ \vert E_b \vert, 
\label{eq:energy}
\end{equation}
where $E$ is the total energy of the 3-$\alpha$ continuum state in the center of mass system and  $E_b$ the energy of the ${}^{12}$C$(2_1^+)$ state with respect to the 3-$\alpha$ threshold energy.
The transition amplitude for the process is given by
\begin{equation}
F^{(\text{B})}(q,\hat{\boldsymbol{x}},\hat{\boldsymbol{y}}) =
\langle \Psi_b \vert H_\gamma \vert   \boldsymbol{q}, \boldsymbol{p} \rangle^{(+)}, 
\end{equation}
where $H_\gamma$ is the electromagnetic transition operator, $\Psi_b$ is the 3-$\alpha$ bound state wave function of  ${}^{12}$C$(2_1^+)$ state, and   $\vert  \boldsymbol{q}, \boldsymbol{p} \rangle^{(+)}$ is the 3-$\alpha$ continuum state initiated by a free 3-$\alpha$ state $\vert \boldsymbol{q},  \boldsymbol{p} \rangle$ with the outgoing boundary condition.

The initial momenta,  $\boldsymbol{q}$ and  $\boldsymbol{p}$, take a variety of values as far as satisfying  the energy conservation relation, 
\begin{equation}
E = \frac{\hbar^2}{m_\alpha} q^2 + \frac{3 \hbar^2}{4m_\alpha} p^2,
\end{equation}
where $m_\alpha$ is the mass of the $\alpha$ particle.
To avoid  a cumbersome procedure  to calculate all $\vert  \boldsymbol{q}, \boldsymbol{p} \rangle^{(+)}$ states, we calculate the inverse reaction of the $3\alpha$ reaction, namely the  E2-photodisintegration of ${}^{12}$C$(2_1^+)$:
\begin{equation}
{}^{12}{\text{C}}(2_1^+) + \gamma \to \alpha + \alpha + \alpha.
\label{eq:photo-dis-C}
\end{equation}
Using the disintegration cross section of this process $\sigma_\gamma(E_\gamma)$,  the $3\alpha$ reaction rate  $\langle \alpha\alpha\alpha \rangle$ at stellar temperature $T$ is calculated  (see, {e.g.},  Ref. \cite{Di10}) by 
\begin{eqnarray}
 \langle \alpha\alpha\alpha \rangle &=& 
 (3)^{3/2} 240 \pi \left( \frac{\hbar}{m_\alpha c} \right)^{3} 
\frac{c}{(k_{\text{B}}T)^3}  e^{-\frac{E_{b}}{k_{\text{B}}T}} 
\cr
&& \times \int_{\vert E_{b} \vert}^\infty  E_\gamma^2
\sigma_\gamma(E_\gamma)  e^{-\frac{E_\gamma}{k_{\text{B}}T}} dE_\gamma,
\label{eq:aaa-sig-g}
\end{eqnarray}
where  $k_{\text{B}}$ is the Boltzmann constant. 
Note that we apply nonrelativistic  kinematics for  the 3-$\alpha$ systems and that we do not consider  a capture to the ${}^{12}$C ground state directly by an electron-positron pair emission in the present work as in the other works \cite{Og09,Ng11,Ng12}.

The three-body disintegration reaction is calculated by defining a wave function \cite{Is94} in an integral equation form, 
\begin{equation}
\vert \Psi \rangle = \frac{1}{E+\imath\epsilon - H_{3\alpha}} H_\gamma \vert \Psi_b \rangle,
\label{eq:Psi_def}
\end{equation}
or in a differential equation form, 
\begin{equation}
\left( E - H_{3\alpha}  \right) \vert \Psi \rangle = H_\gamma \vert \Psi_b \rangle,
\label{eq:Sch-type-eq}
\end{equation}
where $H_{3\alpha}$ is a Hamiltonian of the 3-$\alpha$ system.

Asymptotic form of the wave function evaluated by the saddle-point approximation \cite{Sa77} is  a purely outgoing wave in the three-body space  with the amplitude $F^{(\text{B})}$,
\begin{equation}
\Psi(\boldsymbol{x},\boldsymbol{y}) 
\displaystyle{\mathop{\to}_{\begin{array}{c} x \to \infty \\ y/x~{\text{fixed}}\end{array}} }
\frac{e^{\imath \left( K_0 + {\cal O}(R^{-1}) \right)R }}{R^{5/2}}  
F^{(\text{B})*}(q,\hat{\boldsymbol{x}},\hat{\boldsymbol{y}}),
\label{eq:psi_asym}
\end{equation}
where the hyper radius $R$ and a momentum $K_0$ are given by 
\begin{equation}
R = \sqrt{x^2 + \frac43 y^2}
\end{equation}
and
\begin{equation}
K_0 = \sqrt{ \frac{m_\alpha}{\hbar^2} E},
\end{equation}
$q$ is calculated from the following relation:
\begin{equation}
q =  \frac{1}{\sqrt{ 1+\frac{4}{3}\frac{y^2}{x^2}} } K_0, 
\end{equation}
and  long-range terms due to the Coulomb interaction  \cite{Is09} are expressed just by ${\cal O}(R^{-1})$  for simplicity. 

The photodisintegration cross section is given by the breakup amplitude as
\begin{eqnarray}
\sigma_{\gamma}(E_\gamma) &=& \frac{1}{20\pi} \frac{\hbar}{m_\alpha c} \frac{1}{K_0^3 }  
\left( \frac{3}{4}\right)^{2}
\cr &\times&
\int_0^{K_0} dq q^2 p \vert F^{(\text{B})}(q,\hat{\boldsymbol{x}},\hat{\boldsymbol{y}}) \vert^2. 
\label{eq:sigv}
\end{eqnarray}

We write the 3-$\alpha$ Hamiltonian as
\begin{equation}
H_{3\alpha} = H_0 + \sum_{i=1}^{3} V_{i} + W,
\label{eq:hamiltonian}
\end{equation}
where $H_0$ is the internal kinetic energy operator of the three-body system, 
$V_i$ is a two-body potential (2BP) to describe the interaction between particles $j$ and $k$ consisting of a short-range nuclear potential $V_{i}^{\text{S}}({x}_{i})$ and  the Coulomb potential $V^{\text{C}}(x_i)$ with $Z=2$: 
\begin{equation}
V_i = V^{\text{S}}(x_i) + V^{\text{C}}(x_i) =  V^{\text{S}}(x_i) +  \frac{(Ze)^2}{x_i}, 
\end{equation}
and $W$ is a 3-$\alpha$ potential (3BP). 
Details of potentials used in this work will be described in the next section.

A partial-wave decomposition is performed by introducing an angular function,  
\begin{equation}
\left\vert \theta (\hat{\boldsymbol{x}},\hat{\boldsymbol{y}}) \right)
 = \left[ Y_L(\hat{\boldsymbol{x}}) \otimes Y_\ell(\hat{\boldsymbol{y}}) \right]_{M}^{J},
\end{equation}
where $\boldsymbol{L}$ denotes the relative orbital angular momentum of the pair particles; 
$\boldsymbol{\ell}$ the orbital angular momentum of the spectator particle; 
$\boldsymbol{J} (= \boldsymbol{L}+\boldsymbol{\ell})$ and $M$ the total angular momentum of the three particles and its third component, respectively.
A set of the quantum numbers ($L, \ell, J, M$) is represented by the index $\theta$.

\subsection{Faddeev method}

%
Now, we consider to apply a modified version of the Faddeev three-body method \cite{Fa61} to solve Eq. (\ref{eq:Psi_def}), in which we take into account  the long-range property of the Coulomb ineractions \cite{Sa79}.
Here, we introduce an auxiliary  Coulomb potential $u^{\text{C}}_{i,j}(y_i)$ that acts between the center of mass of the pair $(j,k)$ and the spectator $i$ with respect to the charges of the pair $(i,j)$, 
\begin{equation}
u^{\text{C}}_{i,j}(y_i) = \frac{(Z e)^2}{y_i}.
\label{eq:uhat1} 
\end{equation}
Together with  the similarly defined $u^{\text{C}}_{i,k}(y_i)$, we introduce a Coulomb potential $u^{\text{C}}_i(y_i)$,   
\begin{equation}
u^{\text{C}}_i(y_i) =  u^{\text{C}}_{i,j}(y_i) + u^{\text{C}}_{i,k}(y_i)   = \frac{2 (Z e)^2}{y_i}. 
\label{eq:uhat}
\end{equation}

In the Faddeev theory, a three-body wave function $\Psi$ is decomposed into three (Faddeev) components:
\begin{equation}
\Psi = \Phi^{(1)} + \Phi^{(2)} + \Phi^{(3)}.  
\label{eq:Fad-dec}
\end{equation}
Corresponding to this decomposition,  the three-body potential and  the electromagnetic operator are decomposed into three components:
\begin{equation}
W = W_1 + W_2 +  W_3, 
\label{eq:W-dec}
\end{equation}
and
\begin{equation}
H_{\gamma} = H_{\gamma,1} + H_{\gamma,2} + H_{\gamma,3}     
\label{eq:Hgam-dec}
\end{equation}
with the condition that  $W_i$ and $H_{\gamma,i}$ are symmetric with respect to the exchange of $j$ and $k$.

Modified Faddeev equations \cite{Is94,Sa79} read:
\begin{eqnarray}
\Phi^{(1)} &=& {\cal G}_1(E) H_{\gamma,1} \vert \Psi_b \rangle 
 +{\cal G}_1(E)  \left[ \Delta \Phi \right]^{(1)}, 
\cr
&&
\text{(and ~cyclic~ permutations)},
\label{eq:Fad-eq}
\end{eqnarray}
where the operator ${\cal G}_i(E)$ is a channel Green's function defined as
\begin{equation}
{\cal G}_i (E) \equiv 
\frac1{ E + \imath\varepsilon - H_0 - V_i- u^{\text C}_i}, 
\label{eq:channel-Green}
\end{equation}
and we use a shorthand notation: 
\begin{eqnarray}
 \left[ \Delta \Phi \right]^{(1)}  &\equiv& 
\left( V_1-u^{\text C}_{2,3} \right) \Phi^{(2)} + \left( V_1-u^{\text C}_{3,2} \right) \Phi^{(3)}  
\cr
&+&W_1 \left(\Phi^{(1)} + \Phi^{(2)} + \Phi^{(3)}  \right).
\label{eq:Delta}
\end{eqnarray}

We remark that one obtains the original Schr\"odinger-type equation (\ref{eq:Sch-type-eq}) by summing  up differential equation version of all equations in (\ref{eq:Fad-eq}), and then, using  Eqs.  (\ref{eq:uhat}) -  (\ref{eq:Hgam-dec}). 
We also remark that Eq. (\ref{eq:Fad-eq}) assures that the component $\Phi^{(i)}$ is symmetric under exchange of particles $j$ and $k$, and thus the total wave function $\Psi$, Eq.\ (\ref{eq:Fad-dec}), is totally symmetric with respect to $i$, $j$, and $k$.

Here, we define a set of complete and orthogonal functions describing the angular parts of the three-body system with the state index $\theta$ and the radial part of the spectator particle with momentum $p$, 
\begin{equation}
\left\vert {\cal F}_{{\theta}}(p) \right)
 \equiv \left\vert \theta (\hat{\boldsymbol{x}},\hat{\boldsymbol{y}}) \right) 
 \times 
\sqrt{\frac2{\pi}} \frac{F_\ell[\eta(p), p y]}{y}, 
\label{eq:spectator_F}
\end{equation}
where $F_\ell[\eta(p), p y]$ is the regular Coulomb function: 
\begin{equation}
\left[ T_{\ell}(y) + u^{\text{C}}(y)
\right] F_{\ell}[\eta(p),py] 
   = \left(\frac{3\hbar^2}{4m_\alpha} p^2\right) F_{\ell}[\eta(p),py],
\label{eq:Coulomb-partial}
\end{equation}
with
\begin{equation}
T_{\ell}(y) = -\frac{3\hbar^2}{4m_\alpha} \left(
   \frac{d^2}{dy^2}  - \frac{\ell(\ell+1)}{y^2} \right),
\end{equation}
and a Coulomb parameter $\eta(p) = \frac{2m_\alpha}{3\hbar^2} \frac{2(Ze)^2}{p}$.  

The function $\Phi^{(1)}(\boldsymbol{x},\boldsymbol{y})$ thereby can be expanded as
\begin{equation}
\Phi^{(1)}(\boldsymbol{x},\boldsymbol{y}) =  {\sum_\theta} \int_0^\infty dp
 \left\vert {\cal F}_{{\theta}}(p) \right) 
 \frac{\phi_{{\theta}}(x,p)}{x}, 
\label{eq:phi-xy}
\end{equation}
where the function $\phi_{{\theta}}(x,p)$ is a solution of  an ordinary differential equation:
\begin{equation}
\left[ E_q  - T_{L}(x) - V^{\text S}(x) - V^{\text C}(x)\right] \phi_{{\theta}}(x,p) 
 = {\omega}_{{\theta}}(x,p) 
\label{eq:Sch-Fad}
\end{equation}
with 
\begin{equation}
E_q = \frac{\hbar^2}{m_\alpha} q^2 = E - \frac{3\hbar^2}{4m_\alpha} p^2
\label{eq:Eq}
\end{equation}
and
\begin{equation}
T_L(x) = -\frac{\hbar^2}{m_\alpha} \left(
   \frac{d^2}{dx^2}  - \frac{L(L+1)}{x^2} \right).
\end{equation}
The source function $\omega_{{\theta}}(x,p)$ is given by 
\begin{equation}
 \omega_{{\theta}}(x,p) 
= x \Bigl( {\cal F}_{{\theta}} (p) 
 \left\vert
 H_{\gamma,1}  \Psi_b +  \left[ \Delta \Phi \right]^{(1)}  
\right\rangle.
\label{eq:omega}
\end{equation}

The boundary condition to get a physical solution of Eq. (\ref{eq:Sch-Fad}) depends on $E_q$, and thus on the integral variable $p$ in Eq.\ (\ref{eq:phi-xy}) via Eq. (\ref{eq:Eq}). 
According to the sign of $E_q$, the range of $p$ ($0 \le p < \infty$) is divided into two regions:
(i) $0 \le p \le p_c = \sqrt{\frac{4m_\alpha}{3\hbar^2}E}$, where $E_q\ge0$, and 
(ii)  $p_c < p < \infty$, where $E_q<0$.
Corresponding boundary conditions are
\begin{equation}
\phi_{{\theta}}(x,p) 
  \mathop{\propto}_{x \to \infty}
\left\{
\begin{array}{ll}
u^{(+)}_{L}[\gamma(q),qx] 
& (0 \le p \le p_c),
\\
~
\\
W_{-\gamma(\vert q\vert),L+1/2} ( 2 \vert q \vert x)
& ( p_c < p < \infty),
\end{array}
\right.
\label{eq:bd-cond_3}
\end{equation}
where  
$u^{(\pm)}_L(\gamma,r)$ is defined as 
\begin{equation}
u^{(\pm)}_L(\gamma,r) = 
e^{\mp\imath\sigma_L(\gamma)} 
\left( G_L(\gamma,r) \pm \imath F_L(\gamma,r) \right),
\end{equation}
with $G_{L}(\gamma,r)$ being the irregular Coulomb function,  the factor $\sigma_L(\gamma)$ is the Coulomb phase shift,  $\gamma(q)= \frac{m_\alpha}{2\hbar^2}\frac{(Ze)^2}{q}$, 
and the function $W_{\kappa, \mu}(z)$ is the Whittaker function \cite{Ol10}.
We solve Eq. (\ref{eq:Sch-Fad}) with above conditions by applying usual techniques as in the two-body problem, e.g., the Numerov algorithm \cite{Sa83,Is09}. 

The Faddeev component $\Phi^{(1)}(\boldsymbol{x},\boldsymbol{y})$ has the asymptotic form similar to Eq. (\ref{eq:psi_asym}) with a breakup  amplitude: 
\begin{eqnarray}
f^{(\text{B})}(q,\hat{\boldsymbol{x}},\hat{\boldsymbol{y}}) &=&  e^{\frac{\pi}{4}\imath}
  \left(\frac{4K_0}{3}\right)^{3/2}  \sum_{\theta} 
\left\vert \theta  (\hat{\boldsymbol{x}},\hat{\boldsymbol{y}} )  \right)
\cr
&\times &  
\frac{\imath^{-L-\ell}}{{p}} \frac{m_\alpha/\hbar^2}{1-\imath {\cal K}_{L}({q})}
  \langle \bar{\psi}_{L}({q})\vert
   \omega_{{\theta}}(p) \rangle, 
\end{eqnarray}
where $\bar{\psi}_{L}(x;q)$ is an $\alpha$-$\alpha$  scattering solution with the standing wave boundary condition and ${\cal K}_{L}(q)$ is the scattering $K$-matrix for the two-body scattering (see Appendix C of Ref. \cite{Is09}). 
The total breakup amplitude is thus obtained according to the Faddeev decomposition (\ref{eq:Fad-dec}) as 
\begin{eqnarray}
F^{(\text{B})}(q_1,\hat{\boldsymbol{x}}_1,\hat{\boldsymbol{y}}_1)
&=&
f^{(\text{B})}(q_1,\hat{\boldsymbol{x}}_1,\hat{\boldsymbol{y}}_1)
\cr
&&+
f^{(\text{B})}(q_2,\hat{\boldsymbol{x}}_2,\hat{\boldsymbol{y}}_2)
\cr
&&+
f^{(\text{B})}(q_3,\hat{\boldsymbol{x}}_3,\hat{\boldsymbol{y}}_3).
\end{eqnarray}
%

\section{Calculations}
\label{sec:calculations}

\subsection{Remarks on three-body calculations}

Here, we give some remarks on  3-$\alpha$  calculations.
Some other technical remarks in solving the Faddeev equations for three-body breakup reactions accommodating three-body potentials and  Coulomb potentials are given in Refs. \cite{Sa86,Is87,Is03,Is07,Is09}.

\paragraph{Interactions.}

We use the two-range Gaussian form  \cite{Al66} for the nuclear part of the $\alpha$-$\alpha$ potential,  
\begin{equation}
V^{\text S}(x) =  \hat{P}_{2\alpha,L} V_{R}^{(L)}  e^{-(x/a_R)^2}  + V_{A} e^{-(x/a_A)^2}, 
\label{eq:aa-pot}
\end{equation} 
where $\hat{P}_{2\alpha,L}$ is a projection operator on the $L$ angular momentum $\alpha$-$\alpha$ state.  
In the present work, two different parameter sets will be used: one is from Ref. \cite{Fe96}, which is a slightly  modified version of the model A of the Ali-Bodmer (AB) potential \cite{Al66}, AB(A'); the second set is the model D of the AB potential, AB(D). 
Table \ref{tab:aa-pot-parameters} shows the parameters and calculated properties of $\alpha$-$\alpha$ resonance in comparison with empirical values \cite{Ti04}.  

\begin{table}[t]
\caption{\label{tab:aa-pot-parameters}
Potential parameters of the $\alpha$-$\alpha$ potential, Eq. (\ref{eq:aa-pot}),  
for  AB(A') \cite{Fe96} and the AB(D) \cite{Al66}, and calculated values of the ${}^{8}$Be$(0_1^+)$ resonance energy $E_{r,2\alpha}$ and width $\Gamma_{2\alpha}$.  
Empirical values are taken from Ref. \cite{Ti04}. 
}
\begin{ruledtabular}
\begin{tabular}{cccc} 
Potential & AB(A') & AB(D)  & Empirical \\
\hline
$a_R$ (fm)              & 1.53   & $1/0.70 ~(\sim1.4)$  &\\
$V_{R}^{(0)}$(MeV) & 125.0 & 500.0 &  \\
$V_{R}^{(2)}$(MeV)  &  20.0  &  320.0\\
$a_A$(fm)               & 2.85& $1/0.475~ (\sim2.11)$  \\
$V_A$ (MeV)          &  -30.18 & -130.0 \\ 
$E_{r,2\alpha}$ (keV)            & 93.4     &  95.1    &  91.8  \\
$\Gamma_{2\alpha}$ (eV)     &   8.59   &  8.32  &   $5.57\pm0.25$ \\
\end{tabular}
\end{ruledtabular}
\end{table}

The $\alpha$-$\alpha$ potentials used in this work are shallow, which do not support bound states. 
However, it is known, see, e.g., Refs. \cite{Fi05,Pa08,Su08},  that the use of such shallow $\alpha$-$\alpha$ potentials do not reproduce some 3-$\alpha$ observables, e.g., binding energies and resonance energies.
In order to reproduce these observables, we introduce a 3BP, which depends on the total angular momentum of the 3-$\alpha$ system, which takes a form given in Ref. \cite{Fe96},  
\begin{equation}
V_{3\alpha} = 
\sum_{J=0,2} \hat{P}_{3\alpha,J} W_3^{(J)} \exp\left(-\frac{A_\alpha R^2}{2 b_3^2}\right), 
\label{eq:3bp}
\end{equation}
where $\hat{P}_{3\alpha,J}$ is a projection operator on the 3-$\alpha$ state with the total angular momentum $J$, 
$A_\alpha=m_\alpha/m_N=3.97$ and $b_3=3.9$ fm, and the strength parameters $W_3^{(J)}$ will be determined below. 

\paragraph{Two-body singularity.}

In the integral representation of wave functions, Eq. (\ref{eq:phi-xy}), we need to take care of  the existing of the  ${}^{8}$Be$(0_1^+)$ resonance with the energy $E_{r,2\alpha}$ and the width $\Gamma_{2\alpha}$, which causes a rapid dependence of  $\phi_{{\theta}}(x,p)$ on the variable $p$ through  Eq. (\ref{eq:Eq}). 
As an example, the function $\phi_{{\theta}}(x,p)$ for the inhomogeneous term in Eq. (\ref{eq:Fad-eq}) at  $x= 2.8$ fm and  $E=0.2$ MeV with the AB(D) potential is plotted as a function of $E_q$ instead of $p$.
Here, we set about 30 $p$-mesh (equivalently $E_q$-mesh) points for  $E_{r,2\alpha}- 10 \Gamma_{2\alpha} < E_q < E_{r,2\alpha} + 10 \Gamma_{2\alpha}$. 
The function reveals a sharp $E_q$-dependence around the ${}^{8}$Be$(0_1^+)$ resonance energy, which is safely treated by the condensed mesh points. 
Also, we remark that effects of the function at negative $E_q$ values, which corresponding to closed channel, are significant.  
Thus, in the present calculation, we choose the maximum value of the variable $p$ as the one corresponding to  $E_{p} \approx 160$ MeV. 

\paragraph{Cutoff procedure.}
%
Here, we remark about the introduction of the auxiliary potentials.
Besides the role to introduce the Coulomb distorted spectator function $F_\ell[\eta(p), p y]$, Eq.\ (\ref{eq:Coulomb-partial}), they have another role to play:
In the integral kernel of Eq. (\ref{eq:Fad-eq}),  there appears $u^{C}_{2,3}(y_1)$ with a combination of the Coulomb potential acting particles 2 and 3: 
\begin{equation}
\left( \frac1{x_1} - \frac1{y_2} \right).
\label{eq:x1-y2}
\end{equation}
As explained in Ref. \cite{Is03}, this term is supposed to be a short-range function with respect to the variable $x_1$ because of  a cancellation between two terms, which makes the integral kernel tractable. 
However, while this cancellation holds sufficiently for bound states and continuum states below three-body breakup threshold, it does insufficiently for the case of  the three-body breakup reaction \cite{Is09}. 
To avoid difficulties arising from this, we introduce a mandatory cutoff factor $e^{-(x/R_C)^4}$  to Eq. (\ref{eq:x1-y2}).
This is an approximation made for this calculations. 
To check the  convergence property of the cutoff range $R_C$, we performed calculations with changing  the cut-off radius $R_C$, and found that $R_C=35$ fm is enough to obtain  converged results.

\begin{figure}[t]
\centerline{\includegraphics[width=8.6cm]{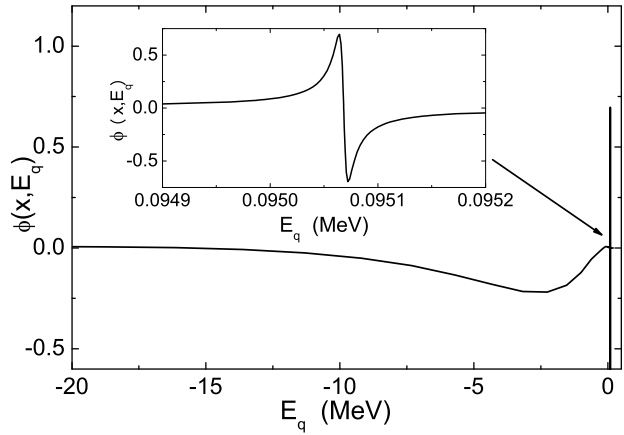}}
\caption{
The function $\phi_{{\theta}}(x,p)$ for the inhomogeneous term in Eq. (\ref{eq:Fad-eq}) at  $x= 2.8$ fm and  $E=0.2$ MeV with the AB(D) potential plotted as a function of $E_q$. 
The insertion is the magnified drawing of $\phi_{{\theta}}(x,p)$ around the 2-$\alpha$ resonance energy.
}
\label{fig:phi-x-p}
\end{figure}

\paragraph{Bound state.}

For the initial  ${}^{12}$C$(2_1^+)$ state, we solve a homogeneous version of Eq. (\ref{eq:Fad-eq}) \cite{Sa86}  taking into account 3-$\alpha$ partial wave states having 2-$\alpha$ states of the angular momentum up to 4 \cite{Fi05,Pa08}. 
The strength parameter of the 3BP for $J=2$ state $W_3^{(2)}$ is determined to reproduce the empirical binding energy of ${}^{12}$C$(2_1^{+})$ state \cite{Aj90}. 
Chosen values of $W_3^{(2)}$  for the  AB(A') and AB(D)  $\alpha$-$\alpha$ potentials are shown in Table \ref{tab:w_3-2-strength}. 

In solving the bound state problem, it is enough to calculate wave functions within a rather restricted area, e.g., ($x \le 12$ fm, $y \le 80$ fm).
However, to use the bound state wave function in solving  Eq. (\ref{eq:Fad-eq}), we need to extend it to large values of the $x$ and $y$ variables.
In actual calculation, we extend the bound state wave function up to  100 fm for both of these variables.
In the present calculations, the extension is performed by expanding the calculated wave function by Gaussian functions. 
The previous results of the present author \cite{Is13,Is12} were insufficient with respect to this expansion, and the present results below are updated, which causes a minor change in results.

\begin{table}[t]
\caption{\label{tab:w_3-2-strength} 
The strength parameters of the three-body potential $W_3^{(2)}$  for the AB(A') and AB(D) $\alpha$-$\alpha$ potential modes together with  calculated values of the binding energy of ${}^{12}$C$(2_1^+)$ state $E_b$. 
The empirical binding energy is taken from Ref. \cite{Aj90}.
}
\begin{ruledtabular}
\begin{tabular}{lcc}
$\alpha$-$\alpha$ model &   $W_3^{(2)}$ (MeV) & $E_b$ (MeV) \\ 
\hline
 AB(A')  &  -56.3 & -2.840 \\
 AB(D)   &  -46.0 & -2.830 \\ 
Empirical &   & -2.8357 \\
\end{tabular}
\end{ruledtabular}
\end{table}

\paragraph{The x- and y-mesh points}

To solve the Sch\"odinger type equation (\ref{eq:Sch-Fad}), the solution $\phi_\theta(x,p)$ is connected to the asymptotic form of Eq. (\ref{eq:bd-cond_3}) at $x=40$ fm in the present calculation.
The function  $\phi_\theta(x,p)$ is then extended up to $x=1000$ fm using the asymptotic form.
Using these functions and Eq. (\ref{eq:phi-xy}), the wave function is extended up to $1000$ fm in the $y$ variable.
These maximum values in $x$ and $y$ variables are checked to give a converged result.

\subsection{Numerical results}

%
For calculations of 3-$\alpha$ continuum states with $J=0$ state, we take into account 3-$\alpha$ partial wave states of  $(L,\ell)=(0,0)$ and $(2,2)$.  
Calculated photodisintegration cross sections reveal a sharp resonance corresponding to the Hoyle state. 
The strength parameter of the 3BP,  $W_3^{(0)}$, is determined to reproduce the empirical resonance energy of the Hoyle state. 
Results for the combination with the AB(A') and for the AB(D) are shown in Table \ref{tab:resonance-parameters}, where  truncated calculations with the $(L,\ell)=(0,0)$ state are denoted by a subscript  ${0}$. 

\begin{figure}[t]
\centerline{\includegraphics[width=8.6cm]{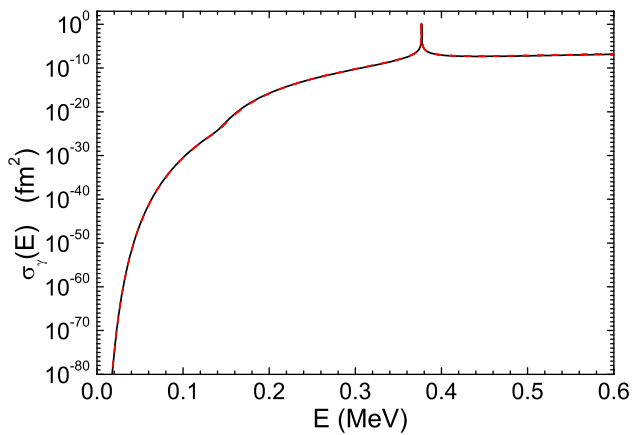} } 
\caption{(Color online) 
Calculated photodisintegration cross section for the process, Eq. (\ref{eq:photo-dis-C}), as a function of the 3-$\alpha$  energy $E$.
The solid line is the results for the AB(A') and the dashed line for the AB(D). 
}
\label{fig:siggamma}
\end{figure}

The partial decay width for the photo-emission process $\Gamma_\gamma$ and the 3-$\alpha$ decay width $\Gamma_{3\alpha}$, which is assumed to equal to the total width, are evaluated by fitting the calculated cross sections around the Hoyle resonance with a Breit-Wigner formula:
\begin{equation}
\sigma_\gamma(E)=\frac{\pi}{10} \left( \frac{\hbar c}{E_\gamma}\right)^2
  \frac{\Gamma_{3\alpha} \Gamma_\gamma}{(E-E_r)^2+\Gamma_{3\alpha}^2/4},
\end{equation}
and are also shown in Table \ref{tab:resonance-parameters}.
Calculated photodisintegration cross sections are plotted in Fig. \ref{fig:siggamma}.

Adapting calculated photodisintegration cross sections to Eq. (\ref{eq:aaa-sig-g}), the $3\alpha$ reaction rates are obtained by numerical integrations. 
The cross sections are normalized to reproduce the empirical value of $\Gamma_\gamma$. 
This is essential to give a reaction rate to agree with that of  the NACRE rate at the resonant region, where the sequential process dominates the reaction and the 3$\alpha$ rate is proportional to  $\Gamma_\gamma$ (See, e.g., Eq. (15) of Ref. \cite{Ga11}).

\begin{table}[t]
\caption{\label{tab:resonance-parameters} 
Strength parameters of the 3BP for 3-$\alpha$ $0^+$ state $W_3^{(0)}$, and calculated resonance parameters of the Hoyle state, $E_{r,3\alpha}$,  $\Gamma_{3\alpha}$, and  $\Gamma_\gamma$ for the Faddeev and CDCC calculations. 
See the text for the subscript $0$ in the Faddeev calculation. 
Empirical values are taken from Ref. \cite{Aj90}.}
\begin{ruledtabular}
\begin{tabular}{lcccc}
Calculation  &$W_3^{(0)}$  & $E_{r,3\alpha}$  & $\Gamma_{3\alpha}$  & $\Gamma_\gamma$ \\
    &  (MeV) & (keV) & (eV) & (meV) \\
\hline
{[Faddeev calculation] } &\\
~~AB(A')   &   -96.2 & 376.966 & 9.1 & 1.8 \\
~~AB(A')$_{0}$   &   -168.0 & 377.929 & 9.5 & 2.7 \\
~~AB(D) & -155.5 & 377.956 & 6.9 & 2.4 \\
~~AB(D)$_{0}$ & -305.5 & 376.724 & 6.4 & 2.9\\
{[CDCC calculation] } &\\
~~AB(A') &  -315.0 & 381.241  &   126 & 4.7 \\
{Empirical}  &     & 379.4 & 8.3$\pm$1.0 & 3.7$\pm$0.5\\
\end{tabular}
\end{ruledtabular}
\end{table}

\begin{figure}[t]
\centerline{
\includegraphics[width=8.6cm]{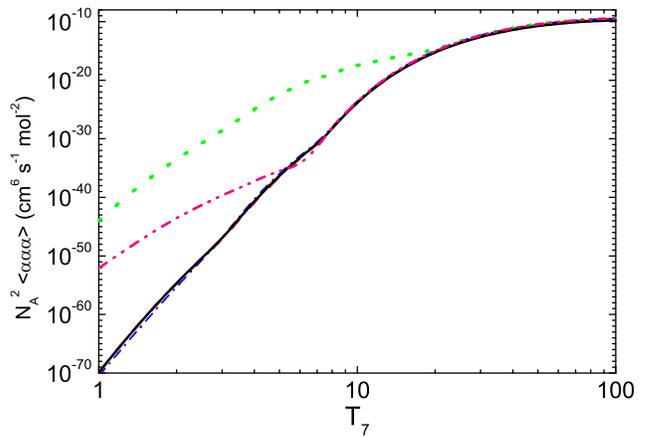}
}
\caption{(Color online) 
The $3\alpha$ reaction rate as a function of the temperature.
The solid line denotes the present calculation for AB(A'); 
the dashed line the AB(D); 
the dot-dashed line the NACRE rate \cite{An99}; 
the dotted line the OKK rate \cite{Og09};
the dot-dot-dashed line the HHR rate \cite{Ng12}. 
}
\label{fig:aaa} 
\end{figure}

\begin{figure}[t]
\centerline{
\includegraphics[width=8.6cm]{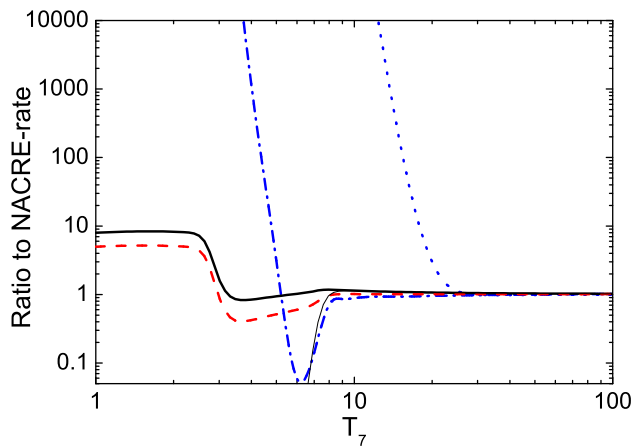}
}
\caption{(Color online)
Ratio of the $3\alpha$ reaction rates to the NACRE rate as a function of the temperature.
The solid and dashed  lines denote the Faddeev calculations for AB(A') and AB(D), respectively; 
the dotted line the OKK rate; 
the dot-dashed line for the HHR rate; 
the thin solid line the Hoyle state contribution for AB(A') (See the text).   
}
\label{fig:aaa-ratio} 
\end{figure}

Calculated $3\alpha$ reaction rates multiplied by the square of  the Avogadro constant $N_{{\text A}}$ by convention, for AB(A') and AB(D) are shown in Fig. \ref{fig:aaa} as a function of the temperature $T_7=T/(10^7 \text{K})$. 
In the figure, reaction rates of the NACRE, OKK, and HHR  are also plotted for comparison. 
In Fig. \ref{fig:aaa-ratio}, ratios of these calculations to the NACRE rate are shown.

Although Table  \ref{tab:resonance-parameters} demonstrates that the determined values of $W_3^{(0)}$ depend on the truncation of the partial wave states,  it turns out that calculated 3$\alpha$ reaction rate essentially do not change once the resonance energy is fitted. 
Actually,  those calculations are indistinguishable even if plotted in Fig. \ref{fig:aaa-ratio}.

Our results of the 3$\alpha$ reaction rate at higher temperatures as $T_7 > 10$ agree with the NACRE rate within a few percents thanks to normalization of the photodisintegration cross section to reproduce the gamma decay width of the Hoyle state. 
However, this contrasts with the calculations of Refs. \cite{Og09,Ng12}, which need to be multiplied by an additional factor after the normalization.  
To see the contribution of the Hoyle state, the 3$\alpha$ rate is calculated by performing the integration of Eq. (\ref{eq:aaa-sig-g})  just around the Hoyle state energy, i.e.,  in the limited range within 10 times of the 3-$\alpha$ decay width.
The result for AB(A') is plotted in Fig. \ref{fig:aaa-ratio} as  thin solid line, which demonstrates that  the reaction rate for $T_7>10$ is actually dominated by  the Hoyle state.

At lower temperatures, the present results are slightly higher than the NACRE rate, which contradicts with the OKK and the HRR rate.
While the present $3\alpha$ rates for AB(A') and AB(D)  are  about $10$ times larger than the NACRE rate at $T_7=1$,  the OKK (HHR) rate is about $10^{26}$ ($10^{18}$) times larger than the NACRE rate at the same temperature. 
These differences will be discussed in the next section.

\begin{figure}[t]
\centerline{
\includegraphics[width=8.6cm]{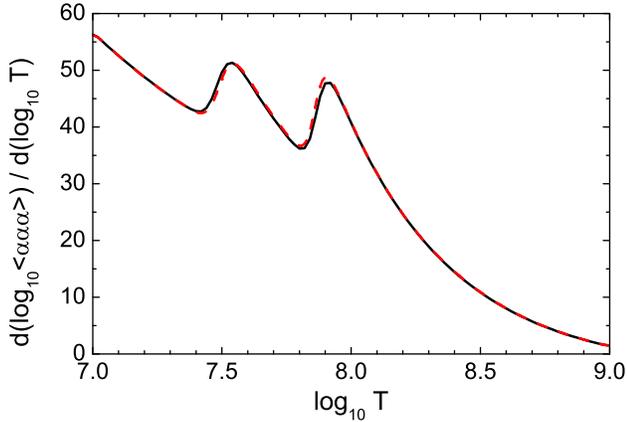}
}
\caption{\label{fig:log_aaa} (Color online)
Temperature-dependence parameter  $d\log_{10}\langle \alpha\alpha\alpha \rangle/ d\log_{10} T$ calculated from the present  $3\alpha$ reaction rate.
The solid and dashed lines denote the Faddeev calculations for AB(A') and AB(D), respectively.
}
\end{figure}

Recently, Suda {\it et al.} \cite{Su11} studied about constraints on the $3\alpha$ reaction rate from a stellar evolution theory. 
Constraints they obtained are:  
(i) $N_{\text A}^2 \langle \alpha\alpha\alpha \rangle < 10^{-29}$ cm$^6$ s$^{-1}$ mol$^{-2}$ at $T\approx10^{7.8} $ K ($T_7 \approx 6.1$); 
(ii) a temperature-dependence parameter $d\log_{10}\langle \alpha\alpha\alpha \rangle/ d\log_{10} T \ge 10$ at $T_7 \approx (10-12)$. 
Fig. \ref{fig:aaa} demonstrates that the present rate satisfies the constraint (i). 
The temperature-dependence parameter calculated from the present result is plotted in Fig. \ref{fig:log_aaa}, which shows that the constraint (ii) is also satisfied for the present rates. 

\section{Discussion}
\label{sec:discussion}

\subsection{CDCC calculation}

In order to discuss the differences between the present Faddeev calculations and the OKK calculation for the 3$\alpha$ reaction rate in some details, we will perform a CDCC calculation for the 3$\alpha$ process.   
However, while the CDCC method was applied to calculate the 3-$\alpha$ continuum states  $\vert  \boldsymbol{q}, \boldsymbol{p} \rangle^{(+)}$ in Ref. \cite{Og09}, it is applied to solve Eq. (\ref{eq:Sch-type-eq}) in the present work.

In the CDCC method \cite{Ka86,Au87}, a three-body wave function is expressed by a particular set of Jacobi coordinates, {e.g.}, $(\boldsymbol{x}_1,  \boldsymbol{y}_1)$, which will be designated as $(\boldsymbol{x},  \boldsymbol{y})$.

We divide the range of the $q$-variable into small intervals of size $\Delta q$, called bin, $[q_{n-1}, q_{n}(=q_{n-1}+\Delta q)]~ (n=1, 2, \dots, N_q)$. 
For each bin, we define a continuum discretized (CD) $\alpha$-$\alpha$ base function by 
\begin{equation}
\hat{\phi}_n(x) = \frac{1}{\sqrt{C_n}} \int_{q_{n-1}}^{q_{n}} dq w_n(q) \phi(x;q),~~~
\end{equation}
where $\phi(x;q)$ is the $\alpha$-$\alpha$ scattering wave functions for the energy $E_q$, 
\begin{equation}
\left[ E_q - T_L(x) - V_1(x) \right] \phi(x;q) = 0,
\end{equation}
$w_n(q)$ is a weight function \cite{Ka86,Au87}, and $C_n$ is the normalization factor,
\begin{equation}
C_n = \int_{q_{n-1}}^{q_{n}} dq \left\vert w_n(q)\right\vert^2. 
\end{equation}

Here, we consider to solve Eq. (\ref{eq:Sch-type-eq})  by expanding the solution by the CD base restricting $L=\ell=L_0=0$ partial wave state,
\begin{equation}
\Psi(\boldsymbol{x},\boldsymbol{y}) = \frac{1}{4\pi} \sum_{n=1}^{N_q}  
\frac{\hat{\phi}_n(x)}{x} \frac{\hat{\psi}_n(y)}{y}, 
\end{equation}
which leads to a set of coupled equations, 
\begin{eqnarray}
&&\sum_{n^\prime=1}^{N_q} \left[
\left( E_{p_n} - T_{\ell}(y)   \right) \delta_{n,n^\prime} - \hat{V}_{n,n^\prime}(y) \right] \hat{\psi}_{n^\prime} (y) 
\nonumber \\
&=& \frac{y}{4\pi}\langle   \hat{\phi}_n \vert H_\gamma \vert \Psi_b \rangle,
\label{eq:cd-eq}
\end{eqnarray}
where
\begin{equation}
E_{p_n} = E - E_{q_n},
\end{equation}
and
\begin{equation}
\hat{V}_{n,n^\prime}(y) = \frac{1}{(4\pi)^2}
 \langle \hat{\phi}_n \vert  V_2 + V_3 + W  
   \vert \hat{\phi}_{n^\prime}  \rangle.
\label{eq:cdcc-pot}
\end{equation}
In calculating this coupling potential, we neglect the angular momentum dependence of the $\alpha$-$\alpha$ potential to avoid any non-locality, and we use the $L=0$ component of the 2BP.

The boundary condition for the function $\hat{\psi}_{n} (y) $ depends on the energy of the spectator particle $E_{p_n}$.
For a positive energy state of the spectator, it is purely outgoing, {e.g.},  
\begin{equation}
\hat{\psi}_{n} (y)
\displaystyle{\mathop{\to}_{y \to \infty}}
 u_0^{(+)}[\eta(p_n), p_n y]  {\cal T}_{n},
\label{eq:cd-bc}
\end{equation}
and then the photodisintegration cross section is given by 
\begin{equation}
\sigma_\gamma = \frac{2}{45\pi} \frac{\hbar c}{mc^2 } 
                         {\sum_{n}}^{\prime} \frac{\left\vert{\cal T}_{n}\right\vert^2}{p_n},
\label{eq:sig_T}
\end{equation}
where the prime means that the summation over $n$ is restricted within a range where $E_{p_n} \ge 0$. 

In the present calculation,  we use 120 averaged states by setting $q_0=0.010$ fm$^{-1}$ ($E_{q_0}=1.0$ keV) with $\Delta q=0.001$ fm$^{-1}$, namely $q_{120}=0.130$ fm$^{-1}$ ($E_{q_{120}}=175$ keV), which is similar choice as the  OKK calculation: 122 states for $q_0=0.008$ fm$^{-1}$ ($E_{q_0}=0.608$ keV) to $q_{122}=0.130$ fm$^{-1}$ ($E_{q_{122}}=176$ keV).
Eq. (\ref{eq:cd-eq}) is integrated up to $y_{\text{max}} = 2500$ fm, and obtained solutions are connected to the outgoing boundary conditions (\ref{eq:cd-bc}). 
In calculating the coupling potential (\ref{eq:cdcc-pot}), the CD-base functions are integrated up to $x_{\text{max}} =5000$ fm.
These maximum values are same as in the OKK calculations.
We use the AB(A') $\alpha$-$\alpha$ potential. 
The same  wave function as in the Faddeev calculation above is used for the initial 3-$\alpha$ ${}^{12}$C$(2_1^+)$ state using the AB(A') model. 

The strength parameter of the 3BP that is determined to reproduce the Hoyle resonance energy is shown in Table \ref{tab:resonance-parameters}.

Due to numerical difficulties in solving Eq. (\ref{eq:cd-eq}) when a channel with negative energy of $E_{p_{n}}$ exists, in the present work, calculations are performed for $E \ge 250$ keV, where all CD channels involved in the calculations are open.

\begin{figure}[t]
\centerline{
\includegraphics[width=8.6cm]{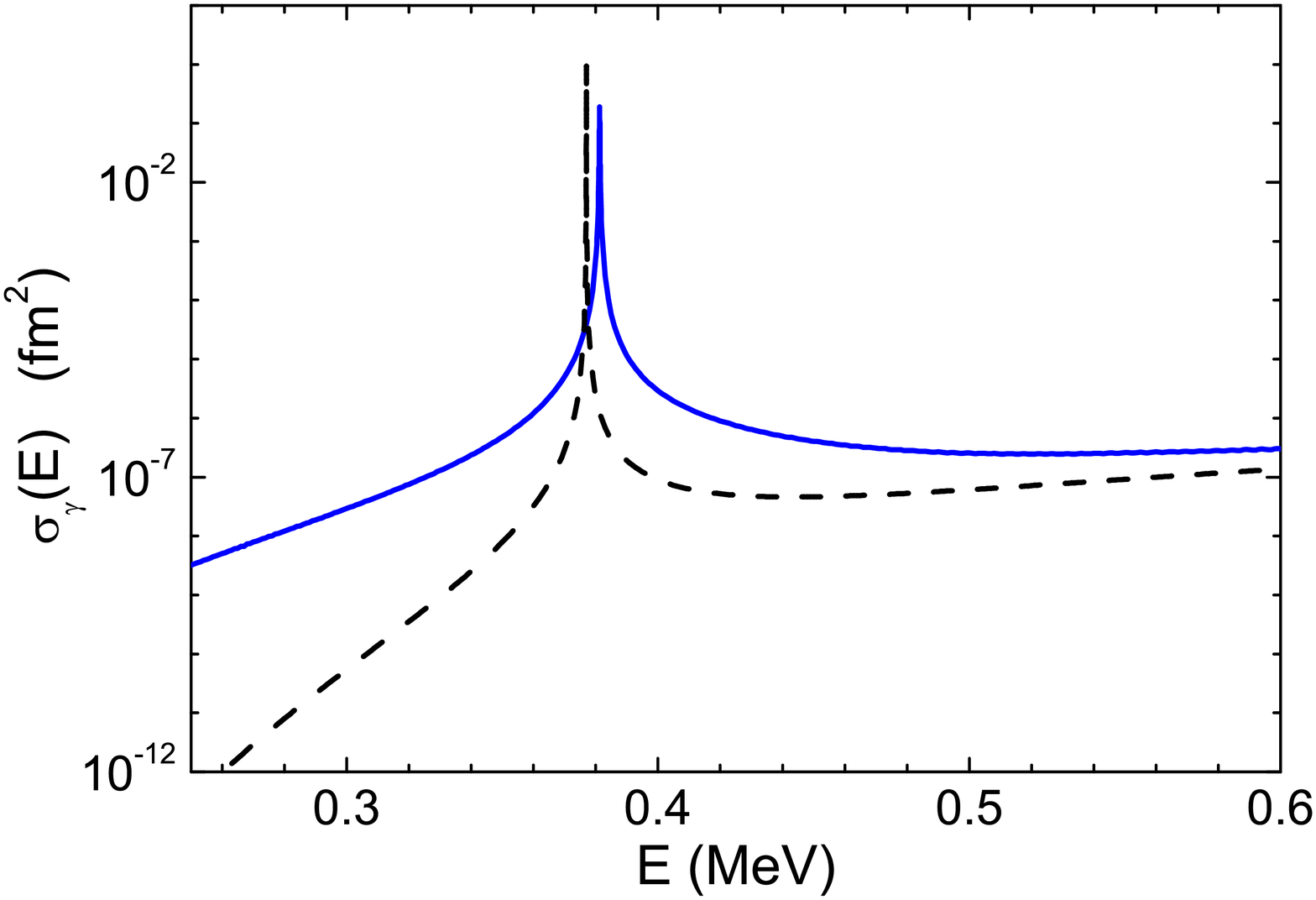}
}
\caption{\label{fig:siggamma-cdcc} 
(Color online) Photodisintegration cross sections of ${}^{12}$C$(2_1^+)$ calculated by the CDCC method (solid line) and the Faddeev method (dashed line). 
}
\end{figure}

In Fig. \ref{fig:siggamma-cdcc}, we plot results of the phododisintegration cross section by the solid line in comparison with the Faddeev result as denoted by the dashed line. 
As is expected, the CDCC cross sections are larger by several orders compared to the Faddeev cross sections.
The resonance parameters calculated by the CDCC method are shown in Table \ref{tab:resonance-parameters},  which shows the calculated width for 3-$\alpha$ decay in the CDCC calculations is 10 times larger than that of the Faddeev calculations and the empirical value.

\begin{figure}[t]
\centerline{
\includegraphics[width=8.6cm]{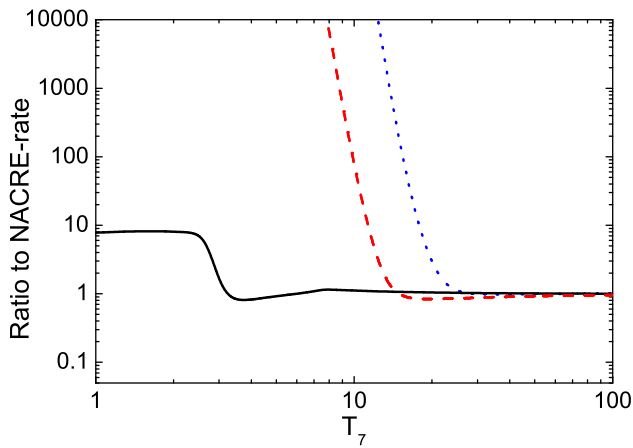}
}
\caption{
(Color online) Ratio of the $3\alpha$ reaction rates to the NACRE rate as a function of the temperature.
The solid   line denotes the Faddeev calculations for AB(A'); 
the dashed line the present CDCC calculation for AB(A');
the dotted line the OKK rate. 
}
\label{fig:aaa-ratio-cdcc} 
\end{figure}

Calculated  3$\alpha$ reaction rate as a ratio to the NACRE rate is plotted in Fig. \ref {fig:aaa-ratio-cdcc}, together with those of the OKK and the AB(A')-Faddeev calculations,  which demonstrates the similar enhancement of the reaction rate as the OKK rate is observed for the present CDCC calculation. 

\subsection{Decay of the Hoyle resonance}

\begin{figure}[t]
\centerline{
\includegraphics[width=8.6cm]{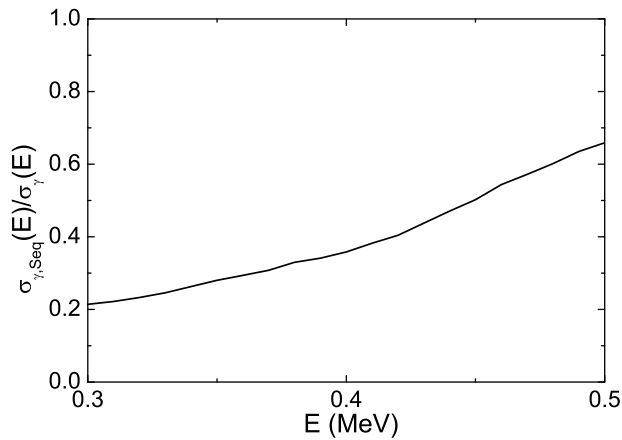}
}
\caption{\label{fig:sigg-resonant-ratio} 
Ratio of the resonant contribution to the phododisintegration cross section of ${}^{12}$C$(2_1^+)$ state calculated by  the  CDCC method for AB(A').
}
\end{figure}

The authors of Ref.\ \cite{Og09} claimed that the significant increase of the OKK rate at low temperatures is due to effects the direct capture reaction, which are enhanced by a proper reduction of the Coulomb barrier between a non-resonant  $\alpha$-$\alpha$ pair and the spectator $\alpha$ particle (see, e.g., Fig. 3 of Ref.\ \cite{Og09}).  
To check the effect of the direct process in the inverse photodisintegration cross section, we extract the sequential cross section as a term of the momentum bin including the $\alpha$-$\alpha$ resonant state from Eq. (\ref{eq:sig_T}), and then define the direct cross section as the rest.    
Fig. \ref{fig:sigg-resonant-ratio} shows the ratio of the sequential cross section to the total cross section of the CDCC calculation for  $0.3$ MeV $ \le E \le 0.5$ MeV. 
The figure shows that the contribution of the sequential cross section accounts for only a small fraction of the total.
This implies a large contribution of the direct cross section and the the reduction of the Coulomb barrier for non-resonant 2-$\alpha$ state as mentioned above. 
In contrast to this, the sequential contribution for the Faddeev calculation defined as an integration around the 2-$\alpha$ resonance energy in Eq. (\ref{eq:sigv}),  turns to contribute more than 99\% of the total cross section. 

Here, we notice that the contribution of the sequential  cross section in the present CDCC calculations becomes only about 30\% of the total even at the Hoyle resonance energy.
This tendency seems to contradict recent experimental results on the decay mechanism of the Hoyle state, which  is produced in different ways:  
by ${}^{40}$Ca + ${}^{12}$C at 25 MeV/nucleon \cite{Ra11},  
by ${}^{10}$C + ${}^{12}$C at 10.7 MeV/nucleon \cite{Ma12}, or 
by ${}^{11}$B$({}^3$He$,d)$ reaction at 8.5 MeV \cite{Ki12b}.
In these experiments, three $\alpha$ particles in the final state are measured in complete kinematics, from which a fraction of   the sequential decay is extracted. 
While Ref. \cite{Ra11} obtained a rather small fraction,  $83(\pm5)$\%, of the sequential decay,  the others \cite{Ma12,Ki12b} obtained the fraction that is almost 100\%.
These results are consistent with the Faddeev calculations, but not with the CDCC calculations.

A possible reason of this difference may be related to an importance of rearrangement channels of the 3$\alpha$ reaction:  
Suppose that a pair of $\alpha$ particles, say 2 and 3,  is in a non-resonant state. 
In the CDCC calculation,  the third $\alpha$ particle 1 feels a rather low Coulomb barrier compared to the case in which the pair is in the ${}^{8}$Be$(0^+_1)$ resonant state as shown in Fig. 3 of Ref.\ \cite{Og09}, and thus the direct reaction proceeds favorably to cause an enhancement of the 3$\alpha$ reaction rate. 
However, in the Faddeev formalism, because of a rearrangement reaction, another pair, say 1 and 3 can form the resonant state, and then the spectator 2 feels a rather high Coulomb barrier, which can suppress the reaction. 
The CDCC calculations do not include such a coupling to rearrangement channels.
As a result, we may say that the direct decay is enhanced for the CDCC calculation due to the lack of rearrangement channels.  

Since the authors of Refs. \cite{Ng11,Ng12} insist that the symmetrization of 3-$\alpha$ wave functions are explicitly took into account in the HHR calculation, the above context may not apply to the difference  between the present calculations and the HHR calculation. 
However, for further studies, it is interesting to see how large is the direct contribution in the HHR calculations.

\section{Summary}
\label{sec:summary}

In this paper, calculations of the $3\alpha$ reaction as a quantum mechanical three-body problem are performed. 
For this, a wave function corresponding to the inverse process is defined and solved by applying the Faddeev three-body theory with accommodating long-range Coulomb force effect. 

Two different models of $\alpha$-$\alpha$ potentials are used supplemented with 3-$\alpha$ potentials to reproduce the binding energy of ${}^{12}$C$(2_1^+)$ state and the resonance energy of the Hoyle state. 
Our results of the $3\alpha$ reaction rate are consistent with the NACRE rate at higher temperatures of $T_7 > 10$, where the sequential process is dominant, and are about $10$ times larger at low temperature of $T_7=1$, although there exists a potential model dependence. 
However, our results contradict recent calculations by the CDCC and HHR methods, which exceeds the NACRE rate by $10^{26}$ and $10^{18}$, respectively,  at $T_7=1$. 

CDCC calculations for the three-body disintegration process are performed, which results similar enhancement of the reaction rate as the OKK rate. 
We found that a remarkable difference between the Faddeev and the CDCC results exist in the contents of decay mode of the Hoyle state: while the sequential decay is dominant for the Faddeev calculation, it is only about 30\% for the CDCC calculation, which contradicts with recent experimental data of the decay of the Hoyle resonance. 


\end{document}